\documentclass{INTERSPEECH2023}

\interspeechcameraready 

\title{Towards multi-task learning of speech and speaker recognition}
\name{Nik Vaessen$^1$, David A. van Leeuwen$^1$}
\address{
  $^1$Institute for Computing and Information Sciences, Radboud University }
\email{nvaessen@science.ru.nl, dvanleeuwen@science.ru.nl}

\def\math#1{\ifmmode#1\else$#1$\fi}

\def\i{{(i)}}
\newcount\exponent
\def\e{\afterassignment\ee\exponent=}
\def\ee{\math{10^{\the\exponent}}}
\def\_{\checkmath_\underscore}
\def\checkmath#1#2{\ifmmode\def\next##1{#1{\rm##1}}\else\let\next=#2\fi\next}

\usepackage{booktabs}
\usepackage{hyperref}

\makeatletter

\renewcommand{\section}{\@startsection
  {section} 
  {1} 
  {0} 
  {-0.5\bigskipamount} 
  {0.5\bigskipamount} 
  {}} 

\renewcommand{\subsection}{\@startsection
  {subsection} 
  {2} 
  {} 
  {\medskipamount} 
  {\medskipamount} 
  {} 
}

\renewcommand{\subsubsection}{\@startsection
  {subsubsection} 
  {3} 
  {} 
  {\smallskipamount} 
  {\smallskipamount} 
  {} 
}

\begin{document}

\maketitle
\begin{abstract}

We study multi-task learning for two orthogonal speech technology tasks: speech and speaker recognition.
We use wav2vec2 as a base architecture with two task-specific output heads.
We experiment with different architectural decisions to mix speaker and speech information in the output sequence as well as different optimization strategies.
Our multi-task learning networks can produce a shared speaker and speech embedding, which on first glance achieve a performance comparable to separate single-task models. 
However, we show that the multi-task networks have strongly degraded performance on out-of-distribution evaluation data compared to the single-task models.
Code and model checkpoints are available at \url{https://github.com/nikvaessen/disjoint-mtl}. 

\end{abstract}


\section{Introduction}

Speech and speaker recognition are, in a sense, orthogonal speech technology tasks. When we develop automatic speech recognition (ASR) systems, a very desirable property is speaker independence: we want the system to perform well irrespective of who uttered the words. Neural ASR models should learn to generate speech embeddings which have minimum variability when the same text is spoken by different speakers. In contrast, when developing speaker recognition (SKR) systems, a very desirable property is text independence: we want the system to perform well irrespective of what was said. Neural SKR models, then, should learn to generate speaker embeddings which have minimum variability when the same speaker utters different texts. We observe a dichotomy where ASR models should be invariant to who speaks while SKR models should be invariant to what is being said. This raises the question: is it possible to train a multi-task learning (MTL) model which can do both speaker and speech recognition, while its components respectively need to be invariant to who is speaking, and what is said?


Besides this interesting academic question, fully-fledged ASR applications often involve speaker recognition components, in order to provide, e.g., speaker-attributed transcriptions, or speaker diarization.  Moreover, in the past, speaker recognition results could be used to improve the performance of ASR models \cite{benzeghiba03_eurospeech,peddinti2015jhu}. Therefore, bringing ASR and SKR together into a single model could reduce the complexity of ASR applications, and has some promise for increased performance. However, we observe the following obstacles in bringing these tasks together:

\begin{enumerate}
    \item Differences in neural architectures for respective tasks, although transformers are bridging this gap.  
    \item Datasets for ASR lack session variability, while datasets for SKR lack transcriptions.
    \item ASR training must be carried out on complete utterances. Typically, ASR datasets do not have aligned transcriptions, while SKR network training is done on short segments as training on long utterances prevents generalization. 
\end{enumerate}

We choose to build on top of the wav2vec2 framework, as the same architecture has been fine-tuned in a single-task learning (STL) setting to both ASR~\cite{Baevski2020}, and speaker recognition~\cite{fan21_interspeech,Vaessen2021}, bridging the gap between neural architectures for ASR and SKR. Our proposed multi-task model is trained with LibriSpeech data for ASR and VoxCeleb for SKR. We train with disjoint steps, meaning batches only contain data from one of the two datasets. This also enables ASR training on complete utterances and SKR training on short segments. This allows us to answer the following research questions:

\begin{enumerate}
    \item Can a transformer-based architecture perform ASR and SKR simultaneously?
    \item Is it feasible to train an MTL model with state-of-the-art datasets for speech recognition and speaker recognition?
    \item Can we train with the complete sentence as input for ASR while using short segments as input for SKR?
\end{enumerate}

\section{Background}

\subsection{Related MTL work}

In~\cite{fan21_interspeech} the wav2vec2 network is used for multi-task learning between the speech tasks of speaker recognition and language identification. 
Their MTL model did not improve on baseline STL performances. 
In~\cite{adi2019reverse} consider whether ASR systems can benefit from MTL learning of speaker recognition, or whether adversarial learning~\cite{ganin2016domain} (AL) is more beneficial. 
Using the WSJ dataset~\cite{paul1992design} and a CNN model, they find similar, but small, improvement gains with MTL and AL. Also, \cite{tang2016multi} train a MTL speech and speaker recognition network on WSJ.
They use two interconnected LSTMs, one for each task. The output of each LSTM is shared in the next time step. 
In~\cite{pironkov2016speaker} an LSTM is trained for ASR, with SKR as auxiliary task, on the TIMIT dataset~\cite{garofolo1993timit}. 
Lastly, the recent Whisper model \cite{radford2022robust} is a multi-task transformer model with impressive ASR performance, which is also capable of doing speech activity detection, language identification and speech translation, but notably, no speaker recognition.

\subsection{Wav2vec2}

An important aspect of the wav2vec2 framework~\cite{Baevski2020} is the application of self-supervised learning to initialize the network weights based on unlabeled data, before fine-tuning the network on (a smaller amount of) labeled data. In this work, we limit ourselves to fine-tuning the network in a multi-task configuration. Further details on the self-supervised learning aspect can be found in the seminal work~\cite{Baevski2020}.

The wav2vec2 architecture consists of three components. First, a 1-d feature extractor CNN processes a raw audio waveform $\mathcal{X} = x_1, \ldots, x_n$ into frames of speech features $\mathcal{Z} = z_1, \ldots, z_m$, with a window size of 20 ms. These features are projected, potentially masked in the time and feature dimension to mimic SpecAugment~\cite{park19e_interspeech} regularisation, and a relative positional embedding is added. The resulting sequence of input vectors, with a receptive field of 2.5\,s, are processed by an encoder network~\cite{devlin2018bert} with multi-head attention transformer layers~\cite{vaswani2017attention} to produce a sequence of output vectors $\mathcal{C}^{L} = c_1^{L}, \ldots, c_m^{L}$, where $L$ specifies the output sequence of a specific transformer layer. The output sequence (of any layer, but usually the last one) can be used by a downstream task. 

For ASR, the output vectors of the wav2vec2 network can represent phones or letters. A single fully-connected (FC) layer can be used to classify each vector, and with CTC loss~\cite{graves2012connectionist} the network is trained end-to-end. For SKR, the output vectors are pooled into a fixed-length speaker embedding~\cite{fan21_interspeech,Vaessen2021}. The network is trained end-to-end by classifying speaker identities using the speaker embedding and a single FC layer. 

\section{Methodology} \label{sec:meth}

\subsection{MTL network architectures}

\subsubsection{Two task-specific heads} \label{sect:vanilla}

Throughout the work we only use the BASE wav2vec2 network architecture with 12 transformer layers. We only make slight modifications for our multi-task purposes by adding two task-specific heads; one for speech recognition, and one for speaker recognition. The automatic speech recognition head consists of a single FC layer which predicts a softmax probability distribution over the vocabulary, for each wav2vec2 output token in the sequence $\mathcal{C}^{12}$. This is equivalent to the original ASR design~\cite{Baevski2020}. 
The speaker recognition head consists of two components. The first part transforms the output sequence into a speaker embedding. The second part, only used during training, is a single FC layer used to classify the train speakers with the speaker embedding. We consider both heads using $\mathcal{C}^{12}$ as input, which implies $C^{12}$ contains speaker and speech information. However, we also experiment with using $C^{n}$ as input for the speaker head instead. In this configuration, the network can gradually remove speaker information from $C^{n+1}$ onward. We chose layer $n=6$ so that half of the network can be solely focused on speech recognition.

\subsubsection{Speaker embeddings}

We compare three strategies to extract a speaker embedding from an output sequence $\mathcal{C}^n$. The first, \textit{mean pooling}, simply aggregates each dimension of the wav2vec2 output vectors $c_1^n, \ldots, c_m^n$ over the time-axis~\cite{fan21_interspeech}. The second, \textit{first pooling}~\cite{Vaessen2021}, does not consider the actual output sequence. Instead, we simply take the first token $c_1^n$ as a speaker embedding. As a third variant, we use the ECAPA-TDNN~\cite{desplanques20_interspeech} architecture to compute a speaker embedding, with $\mathcal{C}^{n}$ as input to ECAPA-TDNN, similar to WavLM~\cite{chen2022wavlm}. Note that by using mean pooling or ECAPA-TDNN, there needs to be speaker information throughout the output sequence, while for first pooling the speech and speaker information can be separated by the transformer layers.

\subsection{Optimization}

We want to train the network on state-of-the-art datasets for speaker and speech recognition. In this section, we suggest two methods for MTL training for speech and speaker recognition. These are based on using  Librispeech~\cite{panayotov2015librispeech}, a well-known dataset for speech recognition, and VoxCeleb~\cite{Nagrani17, Chung18b}, a well-known speaker recognition dataset.


\subsubsection{Disjoint training}

In order to train with LibriSpeech and VoxCeleb, we propose to optimize our network with a disjoint forward step. We assume two datasets, $D_{s}$ and $D_{k}$, base network weights $\theta_{b}$, speech head weights $\theta_{s}$ and speaker head weights $\theta_{k}$. We also have a base network function $N$, a speech recognition head function $H_s$ with loss function $L_s$ as well as a speaker recognition head function $H_k$ with loss function $L_k$.

Each iteration $i$, we sample a speech batch $(x_s^\i, y_s^\i) \in D_{s}$ and a speaker batch $(x_k^\i, y_k^\i) \in D_{k}$. We then apply two forward passes, one on the speech batch, and one on the speaker batch, where we write $p \in \{s,k\}$:
\begin{align*}
    q_p^\i &= N(x_p^\i, \theta_{b}^\i) \\
    \hat{y}_p^\i &= H_p(q_p^\i, \theta_p^\i) \\
    L_p^\i &= L_p(y_p, \hat{y}_p^\i)
\end{align*}

The total loss $L^\i$ is a weighted sum over speech and speaker loss
\begin{equation}
  L^\i = \lambda_s^\i L_s^\i + \lambda_k^\i L_k^\i
\end{equation}
with $\lambda_{s,k}$ the weights for speech and speaker and $\lambda_s+\lambda_k=1$. The gradients for the different parts of the network become
\begin{align}
    \nabla_{\theta_k} L^\i &= \lambda_k \nabla_{\theta_k} L_k^\i \nonumber\\
    \nabla_{\theta_s} L^\i &= \lambda_s \nabla_{\theta_s} L_s^\i \nonumber \\
    \nabla_{\theta_b} L^\i &= \lambda_k \nabla_{\theta_b} L_k^\i + \lambda_s \nabla_{\theta_b} L_s^\i \label{eq:sumgradient}
\end{align}

The weights for the next iteration $\theta_b^{(i+1)}$, $\theta_s^{(i+1)}$ and $\theta_k^{(i+1)}$ are obtained with an optimizer step such as Adam. 



\subsubsection{Joint training}

Most work on MTL assumes training can be done with a `joint' forward step, namely each sample has labels for all tasks. As a baseline, we want to see if training with joint forward steps is effective for MTL of ASR and SKR. We tried two options. The first is to use only data from LibrisSpeech, which has both labels. The second option is to use an ASR model to generate labels for the whole VoxCeleb dataset. We decided to do this with the base\footnote{with \url{https://pypi.org/project/openai-whisper/}}
Whisper \cite{radford2022robust} model. We skip any data labeled as non-English by the Whisper model during training, and normalize the transcript to the character vocabulary of LibriSpeech. 

\subsubsection{Length of audio input during training}

We hypothesize that the discrepancy between audio input lengths for ASR and speaker recognition systems is a potential issue, as the encoder will observe drastically different sequence lengths for each task. We therefore suggest two strategies for cropping the speaker recognition audio segments. The first strategy follows the current paradigm~\cite{Vaessen2021,desplanques20_interspeech,Snyder:2018,lin2020wav2spk,Chung2020MetricSR} and uses crops of 2\,s. The second strategy is to use crops of 10\,s, a value closer to the average length of the audio in LibriSpeech.

\begin{table}[h]
\centering
\caption{Comparison of STL baselines versus joint and disjoint MTL training. Evaluation is done on in-distribution (LS, VOX) and out-of-distribution (HUB5, NIST) data. The second column indicates which training data was used - V2* indicates ASR labels were generated with Whisper. }
\label{tab:mtl-base}
\begin{tabular}{llccccc}
\hline
              &        & \multicolumn{2}{c}{ASR (WER \%)} & \multicolumn{1}{l}{} & \multicolumn{2}{c}{SKR (EER \%)}                   \\ \cline{3-4} \cline{6-7} 
network       & data   & LS-to              & HUB5           & \multicolumn{1}{l}{} & \multicolumn{1}{r}{vox1-h} & \multicolumn{1}{r}{SRE08} \\ \hline
\multicolumn{6}{l}{\textbf{STL}}                                                                           & \multicolumn{1}{r}{}     \\
ASR           & LS     & 10.4            & 40             &                      & -                       & -                        \\
ASR           & V2*    & 16.6            & 25             &                      & -                       & -                        \\
SKR (2s)      & LS     & -               & -              &                      & 33                      & 42                       \\
SKR (2s)      & V2     & -               & -              &                      & 5.1                     & 16                       \\ \hline
\multicolumn{6}{l}{\textbf{MTL (joint, full length samples)}}                                              &                          \\
$\lambda_s=0.5$ & LS     & 15.3            & 48             &                      & 36                      & 40                       \\
$\lambda_s=0.5$ & LS+V2* & 18.1            & 36             &                      & 10.3                    & 24                       \\
$\lambda_s=0.9$ & LS+V2* & 17.5            & 36             &                      & 7.2                     & 26                       \\ \hline
\multicolumn{6}{l}{\textbf{MTL disjoint, 2 sec SKR samples}}                                               &                          \\
$\lambda_s=0.5$ & LS     & 14.5            & 54             &                      & 45                      & 46                       \\
$\lambda_s=0.5$ & LS+V2  & 11.1            & 46             &                      & 41                      & 45                       \\
$\lambda_s=0.9$ & LS+V2  & 11.5            & 48             &                      & 42                      & 46                       \\ \hline
\multicolumn{6}{l}{\textbf{MTL disjoint, 10 sec SKR samples}}                                              &                          \\
$\lambda_s=0.5$ & LS     & 13.6            & 49             &                      & 36                      & 44                       \\
$\lambda_s=0.5$ & LS+V2  & 11.1            & 80             &                      & 4.8                     & 39                       \\
$\lambda_s=0.9$ & LS+V2  & 11.2            & 84             &                      & 4.7                     & 27                       \\ \hline
\end{tabular}
\end{table}

\section{Experiments}

\subsection{Data}

We used the LibriSpeech~\cite{panayotov2015librispeech} (LS) dataset to train and evaluate for speech recognition. The dataset consists of utterances from audio books, read by volunteers. We used all three train subsets, for a total of 960 hours of training data with 2484 speakers. The training audio utterances have a mean of $12.3$ seconds, and a std of $3.84$ seconds. To minimize right-padding (with $0$) in the speech batches, a batch was collected by sampling utterances with similar length. We used the \textit{dev-other} subset to determine a validation word error rate ($\text{WER}\_{val}$). Evaluation was done on the difficult \textit{test-other} (LS-to) subset. The transcriptions were greedily decoded, we did not use a language model. We also create a trial list for dev-other and test-other for SKR. We use all possible pairs, excluding positive trials from the same session (book), and only including same-sex negative trials.

The VoxCeleb1~\cite{Nagrani17} and VoxCeleb2~\cite{Chung18b} (V2) datasets were used to train and evaluate on speaker recognition. The datasets consist of videos of celebrities taken from YouTube. Each speaker has multiple recordings (videos), and each recording has multiple utterances. The VoxCeleb2 ``dev'' subset was used as training, validation, and development data. It has a total of $2305$ hours of data, with 5994 speakers, and a mean utterance length of $7.79$ seconds and a std of $5.22$ seconds. We held-out 194 speakers (97 male/female) to create a development subset. From the remaining 5800 speakers we randomly selected at most two recordings for the validation subset to get a 98\%/2\% train/val split. For the development set we randomly created 100\,k positive and 100\,k negative trial pairs, making sure negative trials are same-sex and positive trials are from 2 different recording sources.
Evaluation was done on the VoxCeleb1 dataset. We used the hard ``VoxCeleb1-H (cleaned)'' trial list (vox1-h). It has 1190 speakers, and each negative trial pair has the same sex and nationality. There is no speaker overlap between VoxCeleb1 and VoxCeleb2. During training and validation, all utterances are randomly cropped to either 2 or 10 seconds. During evaluation, we use the full length of the utterance unless stated otherwise. Trials are scored by computing the cosine similarity between two speaker embeddings, without any further processing. 

To test on out-of-distribution (OOD) data, we also evaluate speech recognition on the English part of HUB5 2000, and speaker recognition on NIST SRE08~\cite{martin2009nist}. For HUB5, we segment the audio based on the ground truth reference to make evaluation easier. We also pre-process the text by removing all annotations and normalizing to the LibriSpeech character vocabulary. For SRE08 we use the 10\,s trials for evaluation. For both datasets we resample the audio to 16Khz.

\begin{table}[]

\centering
\caption{Comparing three methods to extract speaker embeddings from wav2vec2.  Evaluation is done on in-distribution (LS, VOX) and out-of-distribution (HUB5, NIST) data. We vary training with 2s/10s chunks and for MTL also using $\mathcal{C}^6$ or $\mathcal{C}^{12}$.}
\label{tab:skr-var}
\begin{tabular}{lrrrrr}
         & \multicolumn{1}{l}{}   & \multicolumn{1}{l}{}     & \multicolumn{1}{l}{} & \multicolumn{1}{l}{}     & \multicolumn{1}{l}{}     \\
         & \multicolumn{2}{c}{ASR (WER \%)}                  & \multicolumn{1}{l}{} & \multicolumn{2}{c}{SKR (EER \%)}                    \\ \cline{2-3} \cline{5-6} 
SKR head & \multicolumn{1}{c}{LS-to} & \multicolumn{1}{c}{HUB5} & \multicolumn{1}{c}{} & \multicolumn{1}{c}{vox1-h} & \multicolumn{1}{c}{SRE08} \\ \cline{1-3} \cline{5-6} 
\multicolumn{6}{l}{\textbf{STL, $x$/$x$ implies training with 2s/10s SKR samples}}                                                            \\
mean     & -                      & -                        &                      & 5.1/5.1                  & 17/13                    \\
first    & -                      & -                        &                      & 5.4/5.2                  & 19/14                    \\
ECAPA    & -                      & -                        &                      & 6.3/5.8                  & 21/13                    \\ \hline
\multicolumn{6}{l}{\textbf{MTL disjoint, 2 sec SKR samples, $x$/$x$ implies $\mathcal{C}^6$/$\mathcal{C}^{12}$}}                                                          \\
mean     & 13.5/13.4              & 53/52                    &                      & 21/34                    & 40/44                    \\
first    & 13.6/13.9              & 52/53                    &                      & 12/34                    & 29/40                    \\
ECAPA    & 13.2/13.9              & 45/53                    &                      & 9/35                     & 25/39                    \\ \hline
\multicolumn{6}{l}{\textbf{MTL disjoint, 10 sec SKR samples, $x$/$x$ implies $\mathcal{C}^6$/$\mathcal{C}^{12}$}}                                                         \\
mean     & 12.9/12.8              & 51/79                    &                      & 3.9/4.0                  & 31/33                    \\
first    & 13.2/13.4              & 46/79                    &                      & 3.9/4.0                  & 15/16                    \\
ECAPA    & 13.2/12.7                 & 42/83                     &                      & 4.2/4.7                    & 19/16                     \\ \hline
\end{tabular}
\vspace{-4mm}
\end{table}

\subsection{Training protocol}

We use the following training protocol, unless stated otherwise, to balance between spending an equal amount of computational resources on each method, and limiting the required computational budget. 
Each network variant under study is initialized with available\footnote{The pre-trained weights were retrieved from \url{https://huggingface.co/facebook/wav2vec2-base}.}
self-supervised, pre-trained weights~\cite{Baevski2020}, with an identical random seed for all experiments. 
We use a batch size of up to 3.2\,M audio samples ($\leq 200$ seconds) for both tasks~\cite{Baevski2020}.
We use the default regularisation methods for wav2vec2, LayerDrop~\cite{huang2016deep,fan2019reducing} , Dropout~\cite{srivastava2014dropout}, and SpecAugment masking~\cite{park19e_interspeech}. 
The optimizer is Adam~\cite{adamkingma14} and a tri-stage learning rate schedule~\cite{Baevski2020} (10\% warm up, 40\% constant, 50\% exponential decay).
We clip gradients to $[-1, 1]$.
For the first 3\,k steps the whole wav2vec2 network is frozen, only the heads are updated~\cite{Baevski2020}.
The feature extractor CNN is always frozen~\cite{Baevski2020}.
We use CTC loss \cite{graves2012connectionist} as the speech recognition loss, and AAM softmax loss ~\cite{Deng_2019_CVPR,liu19f_interspeech} for the speaker recognition loss with a scale of $30$ and a margin of $0.2$~\cite{desplanques20_interspeech}.
For each network variant we perform a grid search over the learning rates $\{1,3\} \times 10^{-\{4,5,6\}}$ with 200\,k steps.
We stop early if the validation loss has not decreased for 40\,k steps. We validate every 5\,k steps.
For the evaluation, we select an ``optimal'' model and learning rate based on $\frac14 \text{WER}\_{val} + \frac34 \text{EER}\_{val}$. 
Training is done on a machine with a single GPU\footnote{Experiments were done on A5000, A6000 and A100 GPUs.}, 40GB RAM and 12 CPU cores. 
In total 313 days of GPU time was spent on experiments.

\begin{table*}[h]
\centering
\caption{Evaluation of STL and MTL models on cross-disjoint-task and out-of-distribution data. MTL models are trained with $\lambda_s=0.9$.  V2* indicates ASR labels from Whisper during training. For ASR evaluation on Voxceleb we use the vox1-o test set, with labels from Whisper. For SKR evaluation we show results using only the first 2 seconds, or the full utterance of each audio file.}
\label{tab:eval}
\begin{tabular}{llrrrrrrrrrrrr}
          &        & \multicolumn{1}{l}{} & \multicolumn{3}{c}{ASR}                                                     & \multicolumn{1}{l}{} & \multicolumn{3}{c}{SKR (2 sec eval)}                                        & \multicolumn{1}{l}{} & \multicolumn{3}{c}{SKR (full sample eval)}                                  \\ \cline{4-6} \cline{8-10} \cline{12-14} 
model     & data   &                      & \multicolumn{1}{c}{LS-to} & \multicolumn{1}{c}{vox1-o} & \multicolumn{1}{c}{HUB5} & \multicolumn{1}{c}{} & \multicolumn{1}{c}{LS-to} & \multicolumn{1}{c}{vox1-h} & \multicolumn{1}{c}{SRE08} & \multicolumn{1}{l}{} & \multicolumn{1}{c}{LS-to} & \multicolumn{1}{c}{vox1-h} & \multicolumn{1}{c}{SRE08} \\ \hline
STL ASR   & LS     &                      & 10.4                  & 35                   & 40                    &                      & -                      & -                       & -                        &                      & -                      & -                       & -                        \\
STL SKR   & V2     &                      & -                      & -                       & -                        &                      & 4.9                   & 11.0                   & 32                   &                      & 2.2                   & 5.1                    & 16                    \\ \hline
MTL joint & LS+V2* &                      & 17.5                  & 27                   & 36                    &                      & 13.4                  & 21                   & 41                    &                      & 8.5                   & 7.2                    & 26                    \\
MTL DJ 2  & LS+V2  &                      & 11.5                  & 35                   & 48                    &                      & 7.9                   & 12.4                   & 33                    &                      & 40                 & 42                   & 46                 \\
MTL DJ 10 & LS+V2  &                      & 11.2                  & 100                   & 84                    &                      & 44                  & 16                   & 41                    &                      & 42                  & 4.7                    & 27                    \\ \hline
\end{tabular}
\vspace{-4mm}
\end{table*}

\subsection{Comparing MTL optimization strategies}

The first set of experiments are focused on comparing optimization strategies and are shown in Table \ref{tab:mtl-base}. The network architecture in these experiments is fixed; the speech and speaker head both use $\mathcal{C}^{12}$, and the speaker head uses mean pooling. First, observe that single-task training for SKR with LS achieves much worse performance compared to training with V2. It follows that the SKR performance with both joint and disjoint MTL optimization using only LS data is similar. When we do joint optimization with LS and whisper-transcribed V2, the speaker recognition performance drastically improves. Note that MTL training with LS data is worse than STL training with LS data for both SKR and ASR. Looking at disjoint MTL training, we see that using 2\,s SKR chunks during training seemingly leads to no speaker recognition capabilities (discussed further in Section \ref{sec:eval}). Using 10\,s SKR chunks however, makes the MTL outperform the STL baseline on the vox1-h test set, with slightly degraded ASR performance on the LS test set. We also see that the choice of $\lambda_s=0.9$ versus $\lambda_s=0.5$ trades-off SKR and ASR performance.    Lastly, we observe that all MTL models have drastically degraded performance on out-of-distribution test data (Hub5, NIST) compared to the STL baselines. 

\subsection{Varying architectures}

In the second set of experiments we focus on different strategies for extracting speaker information for SKR, and effectively combining it with the speech information for ASR. For all MTL experiments we use $\lambda_s=0.5$ and only train with disjoint steps. We train with either 2 or 10\,s SKR chunks, and place the speaker head at either $\mathcal{C}^{6}$ or $\mathcal{C}^{12}$. The speech head is always at $\mathcal{C}^{12}$. We also apply gradient clipping after summing the gradients instead of before.  Table \ref{tab:skr-var} shows the results. The first observation is that STL speaker recognition actually has better performance when using 10\,s chunks during training, more noticeably on NIST data. Secondly, the specific variant of the speaker head has only a minor effect on the ASR performance. However, using ECAPA-TDNN on $\mathcal{C}^{6}$ seems very effective compared to mean or first pooling. Noticeably, when training with 2\,s second chunks, using $\mathcal{C}^{6}$ instead of $\mathcal{C}^{12}$ seems to result in some SKR capabilities. ASR performance on LS is also worse compared to Table \ref{tab:mtl-base} with equivalent architectures, likely due to changing the clipping strategy.     

\subsection{Different evaluation conditions} \label{sec:eval}

In this section we further analyze the results described in Table \ref{tab:mtl-base}. As we observed decreased performance on out-of-distribution data for the MTL models, we also wanted to observe the performance on cross-disjoint-task data, namely, can we do SKR on LS data, and ASR on V2 data? To evaluate for ASR on V2 we use the transcribed whisper output as ground truth. In Table \ref{tab:eval} we see that this is not always the case. Noticeably, disjoint MTL with 10\,s chunks has a 100\% WER on VoxCeleb data and a 42\% EER on LibriSpeech data. Furthermore, we observed that MTL disjoint training with 2\,s SKR chunks and mean pooling did not show any SKR capabilities. Therefore, perhaps counter-intuitively, we also evaluate on SKR by only using the \emph{first} 2\,s of the utterance, instead of the whole utterance.
\begin{figure}
    \centering
    \includegraphics[width=\columnwidth]{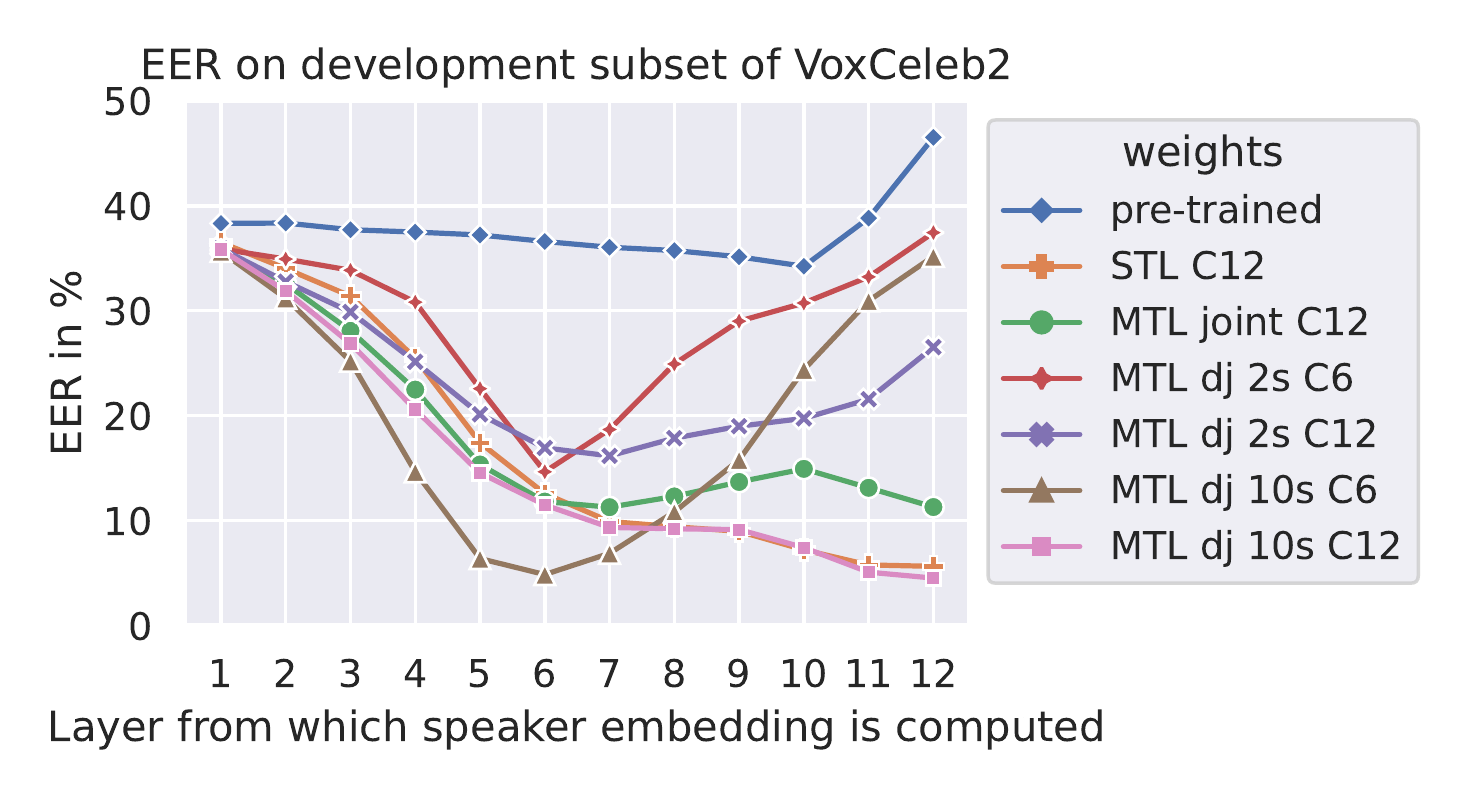}
    \vspace{-8mm}
    \caption{The performance of each wav2vec2 transformer layer by mean pooling the output sequence before (pre-trained, in blue) and after fine-tuning wav2vec2 (STL and 5\,MTL variants). Fine-tuning used speaker head with mean-pooling at~$\mathcal{C}^{6}$ or ~$\mathcal{C}^{12}$.}
    \label{fig:layer_vs_eer}
    \vspace{-8mm}
\end{figure}
We observe that for STL SKR, MTL joint, and MTL disjoint with 10\,s chunks, the SKR performance is worse when using only the first 2\,s of the audio compared to using the full sample. However, MTL disjoint training with 2\,s chunks has decent performance when also evaluating with 2\,s of audio. This compares to no capabilities when evaluating on the full sample. 
Lastly, in Figure \ref{fig:layer_vs_eer} we show how the speaker information is distributed over the network layers. We see that MTL models with a speaker head using $\mathcal{C}^{12}$ actually lose speaker information after $\mathcal{C}^{6}
$, indicating the models attempt to separate speech and speaker information. 

\section{Conclusion}

We have shown that creating an MTL model for speech and speaker recognition is challenging. First, we need multi-labelled data with session variability, LibriSpeech is not sufficient for creating a good SKR model.  Our mitigation strategies with either automatic labels, or disjoint training, have drawbacks. Optimizing a model with disjoint steps doesn't generalize to OOD data. We further saw that MTL models have increased SKR performance, at the cost of decreased ASR performance. It is hard to include speaker information without harming ASR performance. This might be inherent to the MTL loss function, which always needs to trade-off the CTC loss versus the AAM-softmax loss. We believe that future work could focus on integrating speaker information into the CTC loss, by adding e.g., speaker-related targets, and foregoing the need to use two loss functions and two output heads.

\section{Acknowledgements}

This work was sponsored by NWO - Domain Science for the use of supercomputer facilities.

\bibliographystyle{IEEEtran}

\bibliography{main}

\end{document}